\documentclass[12pt]{iopart}

\usepackage{setstack}
\usepackage{iopams}
\eqnobysec

\makeatletter
\def\@fnsymbol#1{^{\thefootnote}\relax}
\makeatother
\newcommand{\be}{\begin{equation}}
\newcommand{\ee}{\end{equation}}

\begin{document}

\title{Ground-state fidelity of Luttinger liquids: A wave functional approach}

\author{John Ove Fj{\ae}restad}
\address{Department of Physics, The University of Queensland, Brisbane, QLD 4072, Australia}
\ead{jof@physics.uq.edu.au}


\begin{abstract}

We use a wave functional approach to calculate the fidelity of ground
states in the Luttinger liquid universality class of one-dimensional gapless
quantum many-body systems. The ground-state wave functionals are discussed
using both the Schr\"{o}dinger (functional differential equation) formulation
and a path integral formulation. The fidelity between Luttinger
liquids with Luttinger parameters $K$ and $K'$ is found to decay exponentially
with system size, and to obey the symmetry $F(K,K')=F(1/K,1/K')$ as a
consequence of a duality in the bosonization description of
Luttinger liquids.

\end{abstract}

\vspace{2pc} \noindent{\it Keywords}: bosonization, Luttinger liquids (theory),
quantum phase transitions (theory).

\maketitle

\section{Introduction}

In recent years it has been realized that concepts from quantum
information theory can be fruitfully applied to analyze and
characterize aspects of the phase diagram of quantum many-body
systems. In particular, the notion of \textit{entanglement} has
proven very powerful \cite{amico-review}. Quantum phase transitions (QPTs)
\cite{sachdev} can be detected by studying ground state entanglement
\cite{amico-review}, and universal terms in the von Neumann entropy
(``entanglement entropy'') have been identified in classes of
critical systems in both one \cite{vidal-etal,cft-EE,disorder-EE}
and two \cite{2d-uni} dimensions, as well as in two-dimensional
topologically ordered phases \cite{top-ent}.

More recently it has also become clear that another useful quantity
for studying QPTs is the \textit{fidelity} between two ground states
corresponding to different parameters in the Hamiltonian. Here the
fidelity is simply (the modulus of) the overlap between the ground
states. The basic idea \cite{zan-paun} is that as one of the
parameter sets is varied so that a quantum phase transition is
crossed, one expects a sharp signature in the fidelity due to the
qualitative difference between ground states in different phases.
This idea has been elucidated and tested on various models, and
generalized and extended in various directions
\cite{zan-paun,zhou-bar,zan-2,cozzini,duncan,oelkers,fidelity-mps,zan-finite-T,buonsante,zan-metric,you,zhou-2,zhou-3,
hamma,campos,chen,gu,bures,mfyang,paunkovic,ning,paunkovic-2,tzeng,zhou-orus-vidal,kwok,zhou-curvature}.

Most of the models which have been investigated from the fidelity
point of view so far have been one-dimensional.
One particularly important universality class in one dimension is
the Luttinger (or Tomonaga-Luttinger) liquid universality class
\cite{haldane}. It includes all critical one-dimensional systems
whose low-energy physics is described by a conformal field theory
with central charge $c=1$, regardless of the details of the
microscopic Hamiltonian and whether it describes fermions, bosons,
or spins \cite{gogetal,giamarchi,giambosons}. The low-energy
effective field theory for Luttinger liquids (LLs) is the Luttinger model
\cite{mattis-lieb}. Recently Yang \cite{mfyang} presented a calculation
of the ground state fidelity of two LLs using the
standard operator formalism.

In this paper we revisit the problem of the fidelity of LL
ground states, using an alternative approach based on
\textit{wave functionals}. In this approach the overlap between
two states $|\Psi_1\rangle$ and $|\Psi_2\rangle$ is expressed
as a functional (or path) integral $\langle \Psi_1|\Psi_2\rangle =
\int {\cal D}\phi\,\Psi_1^*[\phi]\Psi_2[\phi]$,
where $\Psi_1[\phi]$ and $\Psi_2[\phi]$ are wave functionals for
the two states. We note that this expression resembles, and is the field-theoretical
analogue of, the expression for the overlap of two states
$|\psi_1\rangle$ and $|\psi_2\rangle$ in the Schr\"{o}dinger (wave mechanics)
formulation of quantum mechanics,
which is given by an ordinary integral (for simplicity, consider a single
particle in one dimension) $\langle \psi_1|\psi_2\rangle =
\int dx\, \psi_1^*(x)\psi_2(x)$, where $\psi_1(x)$ and $\psi_2(x)$ are
wave functions for the two states.

The ground state wave functional of the LL has been derived by Fradkin \textit{et al.} \cite{fradkin-etal}
and Stone and Fisher \cite{stone-fisher} using path integral methods.
The LL ground state wave functional has also been discussed in the context of the Schr\"{o}dinger
formulation of quantum field theory (the field-theoretical analogue
of the Schr\"{o}dinger formulation of quantum mechanics), in which wave functionals
are obtained as solutions of functional differential equations.\footnote{For
a nice introduction to the Schr\"{o}dinger formulation of quantum field theory,
including comparisons with the operator and path integral
formulations, see \cite{hatfield}.} Fradkin \textit{et al.} \cite{fradkin-etal}
showed that the ground state wave functional they had obtained from their
path integral formulation was indeed the lowest-energy eigen-functional of
the Luttinger model Hamiltonian expressed as a (second-order) functional
differential operator. Closely related Schr\"{o}dinger-type derivations have
been given in \cite{phametal} and \cite{cazalilla}.

In this paper we give an alternative derivation in which the LL
ground state wave functional is obtained as the solution of a (first-order)
functional differential equation that results from translating the
relation $\hat{\beta}_q|\Psi_0\rangle = 0$ to the Schr\"{o}dinger formulation.
Here $|\Psi_0\rangle$ is the LL ground state and $\hat{\beta}_q$
is an arbitrary annihilation operator in the set of canonical boson operators
which diagonalizes the Luttinger model Hamiltonian. We also present an
alternative path integral derivation of the LL ground state wave functional.

In agreement with Yang \cite{mfyang} we find that the ground-state
fidelity of LLs decays exponentially with system size,
but we find that the rate of this exponential decay is a factor of
two smaller\footnote{Yang's result for the fidelity [Equation (10) in \cite{mfyang}] has $\prod_{q\neq 0}$
instead of $\prod_{q>0}$ and is therefore the \textit{square} of our result (\ref{finalfid}).
We note that in Equation (8) for the ground state in \cite{mfyang} the summation should
have been over the positive wavevectors only (or, equivalently, a factor of $1/2$ is missing in the exponent).
When the correct expression for the ground state is used, we find, as expected, that the operator approach
used in \cite{mfyang} gives the same result (\ref{finalfid}) for the fidelity as the wave functional approach.}
than the prediction in \cite{mfyang}.
We stress, however, that this does not change Yang's conclusion \cite{mfyang}
that followed from his application of his fidelity result to the spin-$1/2$
XXZ chain, namely that the \textit{fidelity susceptibility} (the second
derivative of the fidelity \cite{zan-paun,zan-2,cozzini,fidelity-mps,you})
can signal the QPTs in the XXZ chain.\footnote{The $S=1/2$ XXZ spin chain
is in the LL universality class for $-1<\Delta\leq 1$, where
$\Delta=J_{\rm{z}}/J_{\rm{xy}}$ is the exchange anisotropy ratio in the XXZ
model. Yang used his expression for the ground-state fidelity of LLs
to show that the fidelity susceptibility of the XXZ chain signals the QPTs
at $\Delta=\pm 1$ by diverging at those two points. This conclusion is not
affected by the different exponential decay rate of the fidelity found by us,
which only changes the prefactor of the fidelity susceptibility,
not its singular behaviour.} We also find that the ground-state fidelity
of LLs obeys a certain symmetry which we show to be a
consequence of a duality in the bosonization description of LLs.

This paper is organized as follows. In Sec. \ref{gswf} we
present our derivation of the LL ground state
wave functional using the Schr\"{o}dinger formulation. In
Sec. \ref{fidelity} the ground-state fidelity is derived
and its ``duality symmetry'' is explained. Some basic facts
about LLs and their bosonization description,
which form the backdrop for the discussion in the rest of
the paper, are summarized in \ref{LLbasics}. In
\ref{pathintder} a path integral derivation
of the ground state wave functional is presented. In this
paper we follow the bosonization notation of \cite{giamarchi}
rather closely.

\section{Ground state wave functional of the Luttinger liquid: A derivation using the Schr\"{o}dinger formulation}
\label{gswf}

It is well-known that the ground state
wavefunction $\psi_0(x)=\langle x|\psi_0\rangle$ of the harmonic
oscillator can be found from the property $\hat{a}|\psi_0\rangle=0$,
by expressing the bosonic annihilation operator $\hat{a}$ in terms
of the position and momentum operators $\hat{x}$ and $\hat{p}$, and
going to the $|x\rangle$ basis where these operators are represented
as $\hat{x}\to x$, $\hat{p}\to -i\,\partial/\partial x$; this gives
a first-order differential equation for $\psi_0(x)$ which is easily
solved. The derivation that follows is essentially the generalization
of this procedure to the LL. The most important
technical difference from the simple quantum mechanics problem is
that now the argument of the wave ``function'' is a function, not a
number, i.e. we are dealing with a wave function\textit{al}, and
consequently ordinary differentiation is replaced by functional
differentiation.

We start by expanding the operators $\hat{\phi}(x)$ and $\hat{\theta}(x)$
in the Luttinger model Hamiltonian (\ref{HLL}) as
\cite{giamarchi}\footnote{In these expansions we have neglected
the $q=0$ terms (``zero modes'') and also a factor involving a
short-distance cut-off.}
\begin{eqnarray}
\hat{\phi}(x) &=& -\frac{i\pi}{L}\sum_{q\neq 0}
\Bigg(\frac{L|q|}{2\pi}\Bigg)^{1/2}\frac{1}{q}\,e^{- iqx}
(\hat{b}_q^{\dagger}+\hat{b}_{-q}), \label{phiexpand} \\
\hat{\theta}(x) &=& +\frac{i\pi}{L}\sum_{q\neq 0}
\Bigg(\frac{L|q|}{2\pi}\Bigg)^{1/2} \frac{1}{|q|}\,e^{-iqx}
(\hat{b}_q^{\dagger}-\hat{b}_{-q}). \label{thetaexpand}
\end{eqnarray}
The $\hat{b}$-operators obey canonical bosonic commutation relations
$[\hat{b}_q,\hat{b}_{q'}^{\dagger}]=\delta_{q,q'}$, and $L$ is the
length of the system. Next, we make a Bogoliubov transformation to
another set of canonical boson operators $\hat{\beta}_q$,
\begin{equation}
\hat{b}_{q} = \cosh \xi\; \hat{\beta}_{q} - \sinh \xi\;
\hat{\beta}_{-q}^{\dagger}. \label{bogol}
\end{equation}
The parameter $\xi$ is chosen so that the off-diagonal terms in
$\hat{H}$ vanish. The ground state $|\Psi_0\rangle$ is the vacuum of
the $\hat{\beta}$-bosons, i.e. $\hat{\beta}_q|\Psi_0\rangle = 0$ for
all $q\neq 0$, which implies
\begin{equation}
(\hat{b}_q + \tanh \xi\; \hat{b}^{\dagger}_{-q})|\Psi_0\rangle = 0,
\label{gsdefb}
\end{equation}
where
\begin{equation}
\tanh \xi = \frac{1-K}{1+K}. \label{tanhgammaK}
\end{equation}
We invert (\ref{phiexpand})-(\ref{thetaexpand}) to write the
$\hat{b}$-bosons in terms of $\hat{\phi}(x)$ and $\partial_x
\hat{\theta}(x)$,
\begin{equation}
\hat{b}_q = -\frac{i\;\mbox{sgn}(q)}{\sqrt{2\pi L|q|}} \int dx\,
e^{-iqx} \Big[|q|\hat{\phi}(x) + i\,\partial_x \hat{\theta}(x)\Big],
\label{binverse}
\end{equation}
where $\mbox{sgn}(q)$ is the sign of $q$. Inserting (\ref{binverse})
into (\ref{gsdefb}), and using (\ref{tanhgammaK}) and (\ref{momop-phi}),
we get
\begin{equation}
\int dx\, e^{-iqx}\Big[|q|\hat{\phi}(x) +i\pi K \,
\hat{\Pi}_{\phi}(x)\Big]|\Psi_0\rangle = 0.
\label{gseq0}
\end{equation}
We now project this equation onto an eigenstate $|\phi\rangle$ of
$\hat{\phi}(x)$ with eigenvalue $\phi(x)$. Defining the ground state wave functional in
the $\{|\phi\rangle\}$ basis as
\begin{equation}
\Psi_0[\phi]\equiv \langle \phi|\Psi_0\rangle
\label{gswfdefphi}
\end{equation}
and making use of the Schr\"{o}dinger representation of $\hat{\phi}(x)$
and $\hat{\Pi}_{\phi}(x)$ in this basis,
\begin{eqnarray}
\langle \phi|\hat{\phi}(x)|\Psi_0\rangle &=& \phi(x)\Psi_0[\phi], \\
\langle \phi|\hat{\Pi}_{\phi}(x)|\Psi_0\rangle &=& -i \frac{\delta}{\delta
\phi(x)}\Psi_0[\phi],
\end{eqnarray}
we transform (\ref{gseq0}) into the first-order functional
differential equation
\begin{equation}
\int dx\, e^{-iqx}\left[|q|\phi(x) + \pi K \frac{\delta}{\delta
\phi(x)}\right]\Psi_0[\phi] = 0. \label{gseqfuncdiff}
\end{equation}
To solve this equation, we insert the Ansatz\footnote{One can be
led to this Ansatz e.g. by comparing (\ref{gseqfuncdiff})
to the differential equation $(x + x_0^2 \frac{d}{dx})\psi_0(x)=0$
obtained for the simple harmonic oscillator problem discussed at the
beginning of this section, which has the solution
$\psi_0(x)\propto \exp[-(x/x_0)^2/2]$.}
\be
\Psi_0[\phi]\propto \exp\Bigg[-\frac{1}{2\pi K}\int \int dx\,dx'\,\phi(x)g(x-x')\phi(x')\Bigg].
\label{gswf-phi}
\ee
Here the coefficient matrix $g(x,x')$ in the quadratic form was taken to be symmetric
without loss of generality, and also translationally invariant, i.e.
$g(x,x')=g(x-x')$. Calculating the functional derivative in
(\ref{gseqfuncdiff}) and introducing the Fourier transforms $\tilde{\phi}(q)$ and
$\tilde{g}(q)$ of $\phi(x)$ and $g(x)$, respectively, we find
$[|q|-\tilde{g}(q)]\tilde{\phi}(q)=0$, i.e.,
\be
\tilde{g}(q)=|q|.
\label{gq}
\ee
From this result $g(x)$ can be found; however, since an explicit expression for $g(x)$
will not be needed in the following, we relegate a discussion of it to the path integral
derivation of $\Psi_0[\phi]$ in \ref{pathintder} where it comes up naturally.

In section \ref{fidelity} the fidelity will be calculated from $\Psi_0[\phi]$. For the purpose
of understanding a symmetry that the fidelity will be shown to possess, we will now briefly
also discuss the ground-state
wave functional in the basis in which the operator $\hat{\theta}(x)$ is diagonal. This
wave functional, defined as $\bar{\Psi}_0[\theta]\equiv \langle \theta|\Psi_0\rangle$
where $|\theta\rangle$ is an eigenstate of $\hat{\theta}(x)$ with eigenvalue $\theta(x)$,
can e.g. be derived in a way that is completely analogous to the derivation of $\Psi_0[\phi]$
above. Expressing the $\hat{b}$-operators in terms of $\hat{\theta}(x)$ and $\partial_x \hat{\phi}(x)$,
i.e. $\hat{b}_q = i(2\pi L|q|)^{-1/2}\int dx\,e^{-iqx}[|q|\hat{\theta}(x)+i\partial_x \hat{\phi}(x)]$,
(\ref{gsdefb}) leads to an equation that is identical in form to (\ref{gseq0}) but differs by the replacements
$\phi\to \theta$ [which here amounts to $\hat{\phi}(x)\to \hat{\theta}(x)$ and
$\hat{\Pi}_{\phi}(x)\to \hat{\Pi}_{\theta}(x)$] and $K\to 1/K$. It follows that $\bar{\Psi}_0[\theta]$
can be obtained from $\Psi_0[\phi]$ by making the same replacements. Thus
\be
\bar{\Psi}_0[\theta] \propto \exp\Bigg[-\frac{K}{2\pi}\int \int dx\,dx'\,\theta(x)g(x-x')\theta(x')\Bigg].
\label{gswf-theta}
\ee
The property that $K\leftrightarrow 1/K$ when $\phi\leftrightarrow \theta$ holds more generally
\cite{gogetal,giamarchi} and is referred to as a duality; $\phi(x)$ and $\theta(x)$ are often
referred to as dual fields. Thus one can regard the wavefunctionals $\Psi_0[\phi]$ and $\bar{\Psi}_0[\theta]$
as dual representations of the LL ground state $|\Psi_0\rangle$.

\section{Fidelity between Luttinger liquid ground states}
\label{fidelity}

In this section we discuss the ground state fidelity between two
LLs with Luttinger parameters $K$ and $K'$.\footnote{The
Luttinger velocities $u$ and $u'$ do not come into consideration here since the
LL ground states are independent of these velocities.}
Denoting the two (normalized) ground states by
$|\Psi_{0,K}\rangle$ and $|\Psi_{0,K'}\rangle$, the fidelity between
them is defined as the modulus of their overlap,
\be
F(K,K')=|\langle \Psi_{0,K}|\Psi_{0,K'}\rangle|.
\label{fidelitydef}
\ee

We begin by considering, as in the previous section, a continuum
field theory description of a system of length $L$ with periodic boundary conditions. Using the
resolution of the identity in the form $I=\int \prod_x d\phi(x)\,|\phi\rangle\langle \phi|$,
the ground state overlap can be expressed as a path integral involving the ground state wave functionals
in the $\{|\phi\rangle\}$ basis,
\begin{eqnarray}
\lefteqn{\langle \Psi_{0,K}|\Psi_{0,K'}\rangle = \int \prod_x d\phi(x)\;
\Psi_{0,K}^*[\phi]\Psi_{0,K'}[\phi]} \nonumber \\ & & \fl
=\,{\cal N}_K\, {\cal N}_{K'} \int \prod_x d\phi(x)\,
\exp\Bigg[-\frac{1}{2\pi}\Bigg(\frac{1}{K}+\frac{1}{K'}\Bigg)\int\int dx\, dx'\, \phi(x) g(x-x') \phi(x')\Bigg].
\label{overlap0}
\end{eqnarray}
Here ${\cal N}_K$ is the normalization factor in $\Psi_{0,K}[\phi]$.
The quadratic form is diagonalized by introducing the Fourier transform
$\phi(x)=(1/L)\sum_q \tilde{\phi}(q)e^{iqx}$ and similarly for $g(x)$;
the $q$'s are discrete (due to the finite size $L$) and unbounded (due to
the continuum nature of the theory). This gives
\begin{eqnarray}
\lefteqn{\hspace{-0.5cm}\langle \Psi_{0,K}|\Psi_{0,K'}\rangle =
{\cal N}_K\,{\cal N}_{K'}\, J\int d\tilde{\phi}(0)\int
\prod_{q>0}d\tilde{\phi}^*(q)d\tilde{\phi}(q)} \nonumber \\ & & \fl
\times \exp\Bigg[-\frac{1}{\pi}\Bigg(\frac{1}{K}+\frac{1}{K'}\Bigg)\frac{1}{L}\sum_{q>0}
\tilde{g}(q)|\tilde{\phi}(q)|^2\Bigg] = {\cal
N}_K {\cal N}_{K'} J \Omega_0 \prod_{q>0} \frac{2\pi
i}{\frac{1}{\pi}\big(\frac{1}{K}+\frac{1}{K'} \big)\frac{1}{L}\tilde{g}(q)}.
\label{overlap}
\end{eqnarray}
Here $J$ is the Jacobian for the change of integration variables
from $\phi(x)$ to $\tilde{\phi}(q)$ in the path integral measure,
and $\Omega_0=\int d\tilde{\phi}(0)$ is a (divergent) quantity which
arises because $\tilde{g}(0)=0$. Setting $K=K'$ in (\ref{overlap}) and
using $\langle \Psi_{0,K}|\Psi_{0,K}\rangle = 1$ we find ${\cal
N}_K=\left(J\Omega_0 \prod_{q>0}\frac{i\pi^2
KL}{\tilde{g}(q)}\right)^{-1/2}$. Inserting this in (\ref{overlap}), most
quantities cancel out, resulting in the following expression for the
ground state fidelity of two LLs \cite{mfyang} (see also footnote 2):
\be
F(K,K')=\prod_{q>0}\frac{2}{\sqrt{\frac{K}{K'}}+\sqrt{\frac{K'}{K}}}.
\label{finalfid}
\ee
As it stands, the rhs of this expression vanishes (for $K\neq K'$) even for a system
of finite size $L$, because we haven't yet taken into account the short-distance
cut-off of the field theory. Thus it is necessary to regularize (\ref{finalfid}). In the
present context this is most straightforwardly done by considering its logarithm which involves the sum $\sum_{q>0}1$.
We use a \textit{soft} cut-off $\alpha$ to remove this divergence by inserting a factor $e^{-\alpha q}$
in the sum. For $L\gg \alpha$ this then gives
$F(K,K')=[\kappa(K,K')]^{L/2\pi \alpha}$, where we have defined $\kappa(K,K')=2/(\sqrt{K/K'}+\sqrt{K'/K})$.
As $1/\alpha$ is a measure of the (effective) maximum wavevector $q_{\rm{max}}$, and the distance between adjacent
wavevectors is $\Delta q = 2\pi/L$, the exponent $L/2\pi \alpha$ appearing in
the fidelity is $\sim q_{\rm{max}}/\Delta q$ and thus just represents the effective
number of wavevectors in the product in (\ref{finalfid}) after the regularization, as
expected.

An alternative regularization approach involves discretizing the field theory, i.e.,
putting it on a one-dimensional lattice with lattice constant $a$, so that the number of sites is $N=L/a$.
Here the lattice constant plays the role of a \textit{hard} cut-off. The $q$'s then become restricted
to the first Brillouin zone, i.e., $|q|< \pi/a \equiv q_{\rm{max}}$, and it can be shown that (\ref{gq}) must
be replaced by $\tilde{g}(q)=a^{-1}\sqrt{2(1-\cos qa)}$ (which reduces to (\ref{gq}) in the limit $qa\to 0$).
Calculating the fidelity, one again arrives at the result (\ref{finalfid}) where now the upper limit
$\pi/a$ for the $q$'s is implicitly understood. For $N\gg 1$ this gives $F(K,K')=[\kappa(K,K')]^{N/2}$.
The obtained exponent ($N/2$) can be written $q_{\rm{max}}/\Delta q$, consistent with what
we found when using a soft cut-off. Thus for $K\neq K'$ the fidelity is seen to decay exponentially with system
size.

From these results we can deduce the fidelity per site $d(K,K')$ (equivalently the fidelity per wavevector),
defined as \cite{zhou-bar} $\ln d(K,K')=N^{-1}\ln F(K,K')$. This gives
\be
d(K,K')=\sqrt{\frac{2}{\sqrt{\frac{K}{K'}}+\sqrt{\frac{K'}{K}}}}.
\ee
It can be argued (cf. \cite{zhou-bar}) that this is a more fundamental quantity than the fidelity itself,
because unlike $F(K,K')$, $d(K,K')$ is independent of the short-distance cut-off and remains finite in the
thermodynamic/continuum limit $N=L/a\to \infty$ where $F(K,K')$ vanishes.

For the exactly solvable Luttinger liquid field theory studied here, both regularization approaches
considered above are viable for studying the fidelity. The first approach is closer in spirit to most
of the literature on Luttinger liquids. On the other hand, the second approach, based on lattice
regularization, is more suitable for numerical studies of the fidelity of more general continuum
field theories that are not analytically solvable \cite{huan-private}. Again, the main quantity of interest
is the fidelity per site, which can be numerically extracted from the fidelity of the discretized
version of the field theory in the limit as the lattice constant is sent to zero.
As such numerical studies could also address field theories describing gapped (massive)
phases, they would open up the possibility of investigating quantum phase transitions in continuum field
theories by studying the behaviour of the fidelity per site as a function of the coupling constants.

We conclude this section by discussing some aspects of the fidelity's dependence
on the Luttinger parameters. We first note that (\ref{finalfid}) satisfies $F(K,K)=1$ and
$F(K,K')=F(K',K)$. These two properties are however evident already in the definition
(\ref{fidelitydef}) of the fidelity. A much more interesting property of
(\ref{finalfid}) is the symmetry
\be
F(K,K')=F(1/K,1/K').
\label{fidduality}
\ee
As we will now show, this symmetry is a consequence of the duality
of the LL discussed at the end of section \ref{gswf}.
Expressing the overlap $\langle \Psi_{0,K}|\Psi_{0,K'}\rangle$ in
$F(K,K')$ in terms of the $\{|\theta\rangle\}$ basis gives
(cf. the first line in (\ref{overlap0}))
\be
\langle \Psi_{0,K}|\Psi_{0,K'}\rangle = \int \prod_x d\theta(x)\;
\bar{\Psi}^*_{0,K}[\theta]\bar{\Psi}_{0,K'}[\theta].
\label{fidtheta}
\ee
Inserting (\ref{gswf-theta}) one sees that the resulting expression
differs from the second line of (\ref{overlap0}) only by the replacement
$(K,K')\to (1/K,1/K')$ which thus must leave the overlap and hence
$F(K,K')$ invariant. Thus (\ref{fidduality}) follows.\footnote{We note that in the
alternative operator approach, which was used to calculate the fidelity
in \cite{mfyang}, the symmetry (\ref{fidduality}) can be
understood as follows. If $K\to 1/K$, (\ref{tanhgammaK}) gives
$\tanh \xi \to -\tanh \xi$, i.e. $\xi\to -\xi$. The fidelity only
depends on $\cosh(\xi-\xi')$ \cite{mfyang} and is therefore
invariant under $(\xi,\xi')\to(-\xi,-\xi')$.}

\ack

We thank Andrew Doherty and Huan-Qiang Zhou for discussions, and
Ross McKenzie and Min-Fong Yang for comments. This work was
supported by the Australian Research Council.


\appendix

\section{Some basics of Luttinger liquids and their bosonization description}
\label{LLbasics}

The effective low-energy Hamiltonian for a LL is given
by the Luttinger model. Using the bosonization description
\cite{gogetal,giamarchi}, the Luttinger model Hamiltonian can be
written
\begin{equation}
\hat{H} = \frac{u}{2\pi}\int dx\;\Bigg[K:(\partial_x
\hat{\theta}(x))^2: + \frac{1}{K}:(\partial_x \hat{\phi}(x))^2:
\Bigg]. \label{HLL}
\end{equation}
Here $\hat{\phi}(x)$ and $\hat{\theta}(x)$ are Hermitian fields and
$: \ldots :$ represents normal-ordering. The operators
\begin{eqnarray}
\hat{\Pi}_{\phi}(x) &\equiv & \frac{1}{\pi}\partial_x \hat{\theta}(x), \label{momop-phi} \\
\hat{\Pi}_{\theta}(x) &\equiv & \frac{1}{\pi}\partial_x \hat{\phi}(x), \label{momop-theta}
\end{eqnarray}
are the conjugate momenta of $\hat{\phi}(x)$ and $\hat{\theta}(x)$, respectively,
i.e. the following canonical equal-time commutation relations hold:
\begin{eqnarray}
[\hat{\phi}(x),\hat{\Pi}_{\phi}(x')] &=& i\delta(x-x'), \label{commrel-phi} \\
\mbox{}[\hat{\theta}(x),\hat{\Pi}_{\theta}(x')] &=& i\delta(x-x'). \label{commrel-theta}
\end{eqnarray}

The LL is characterized by the two parameters $u$ and
$K$ \cite{haldane,gogetal,giamarchi}. The former is the velocity of the low-energy excitations (whose
energy disperses linearly with the wavevector), while the latter,
known as the Luttinger parameter, determines the exponents of the
asymptotic power-law decays of the correlation functions of the
LL. The dependence of $K$ and $u$ on the parameters in a lattice model in the
LL universality class may be determined numerically or,
in some cases, analytically. Analytical expressions
are usually limited to regions of parameter space which are within
reach of approximate analytical treatments, unless the model in
question is integrable, in which case exact analytical
results for $K$ and $u$ valid in the entire LL
regime may be available. A model of this latter type is
the spin-$1/2$ XXZ chain, which is exactly solvable by the Bethe
Ansatz \cite{johnson}; the LL regime for this model
is $-1<\Delta\leq 1$ where $\Delta=J_{\rm{z}}/J_{\rm{xy}}$ is the exchange
anisotropy. By comparing LL predictions with
the exact solution one finds \cite{lutt-xxz} $K=\pi/[2(\pi-\arccos \Delta)]$
and $u=\pi J_{\rm{xy}} \sqrt{1-\Delta^2}/(2\arccos \Delta)$.
Thus as $\Delta$ is reduced from 1, $K$ increases from $1/2$,
passing through $1$ at $\Delta=0$ and diverging as $\Delta\to -1$.

\section{A path integral derivation of the ground state wave functional}
\label{pathintder}

In this Appendix we present a path integral derivation of the LL ground
state wave functional $\Psi_0[\phi]$. Like the path integral derivation of
this quantity given in \cite{stone-fisher} (see also \cite{lw}), the derivation
discussed here is based on a path integral representation of the matrix element
$\langle \phi_f|e^{-\hat{H}\Delta \tau}|\phi_i\rangle$ of the
imaginary-time evolution operator of the Luttinger model in the limit
$\Delta \tau\to \infty$. In contrast to
\cite{stone-fisher} and \cite{lw}, however, we do not set $\phi_i(x)=0$,
i.e. we take both $\phi_f(x)$ and $\phi_i(x)$ to be arbitrary. Furthermore,
we determine the dependence of this matrix element on $\phi_f(x)$ and $\phi_i(x)$
for an arbitrary time interval $\Delta \tau$. Another difference from \cite{stone-fisher}
is that we use Poisson's integral formula instead of
Green function methods to find the classical action.

Expanding $|\phi_i\rangle$ and $|\phi_f\rangle$ in terms of the complete set
of eigenstates $\{|\Psi_n\rangle\}$ of $\hat{H}$ gives
\be
\langle \phi_f|e^{-\hat{H}\Delta \tau}|\phi_i\rangle =
\sum_{n} \Psi_n[\phi_f]\Psi_n^*[\phi_i]e^{-E_n \Delta \tau},
\label{me-series}
\ee
where $\{E_n\}$ is the corresponding set of eigenvalues and $\Psi_n[\phi]\equiv\langle \phi|\Psi_n\rangle$. In the limit
$\Delta \tau\to \infty$ the contribution from the excited states in the sum is completely suppressed compared to
that of the ground state, giving
\be
\Psi_0[\phi_f]\Psi_0^*[\phi_i]=\lim_{\Delta \tau\to \infty}\frac{\langle \phi_f|
e^{-\hat{H}\Delta \tau}|\phi_i\rangle}
{e^{-E_0 \Delta \tau}},
\label{gswf-from-me}
\ee
where $E_0$ is the ground state energy. This relation will be used to deduce the ground state wave functional
$\Psi_0[\phi]$.

A path integral representation of $\langle \phi_f|e^{-\hat{H}\Delta \tau}|\phi_i\rangle$
can be obtained by Trotter-decomposing
the exponential and inserting resolutions of the identity in a standard
way (see e.g. \cite{greiner}). This gives
\begin{equation}
\langle \phi_f|e^{-\hat{H}\Delta \tau}|\phi_i\rangle \propto
\int^{\phi(x,\tau_f)=\phi_f(x)}_{\phi(x,\tau_i)=\phi_i(x)}
{\cal D}\phi(x,\tau)\;\exp{(-S[\phi(x,\tau)])},
\label{gswf-phi-PI}
\end{equation}
where the Euclidean action is
\begin{equation}
S[\phi(x,\tau)] = \frac{1}{2\pi K}\int dx \int_{\tau_i}^{\tau_f} d\tau\;
\left(u (\partial_x
\phi)^2+\frac{1}{u}(\partial_{\tau}\phi)^2\right).
\label{actionT=0}
\end{equation}
In these expressions $\tau_i$ and $\tau_f$ are the initial and final time, respectively, with
$\Delta \tau = \tau_f-\tau_i$. Making the variable change $y = u\tau$, the action takes the more
symmetric form $S[\phi(x,y)]=(2\pi K)^{-1}\int dx \int_{y_i}^{y_f} dy\;[(\partial_x \phi)^2 +
(\partial_y \phi)^2]$
where $y_i=u\tau_i$, $y_f=u\tau_f$. We see that
$\langle \phi_f|e^{-\hat{H}\Delta \tau}|\phi_i\rangle $ is given by a path integral of
$\exp(-S[\phi(x,y)])$ over real-valued functions $\phi(x,y)$ defined on the horizontal strip in the $xy$ plane
bounded by $y=y_i$ and $y=y_f$ , with boundary conditions $\phi(x,y_i)=\phi_i(x)$ and
$\phi(x,y_f)=\phi_f(x)$
on the lower and upper edge of the strip, respectively.

Because the action is quadratic, the path integral can be calculated
exactly. We expand $S$ around the classical action $S_{\rm{cl}}$
corresponding to $\phi_{\rm{cl}}(x,y)$, the solution of the
classical equation of motion $\delta S/\delta \phi(x,y)=0$ (which
for the LL is the Laplace equation in two dimensions),
subject to the boundary conditions. The path integral in
(\ref{gswf-phi-PI}) can then be written as a product of
$\exp{(-S_{\rm{cl}})}$ and a path integral over the deviations
$\phi(x,y)-\phi_{\rm{cl}}(x,y)$ from the classical solution. The
boundary conditions at $y=y_i$ and $y=y_f$ imply that the entire
dependence on $\phi_i(x)$ and $\phi_f(x)$ lies in $S_{\rm{cl}}$, i.e.
\be
\langle \phi_f|e^{-\hat{H}\Delta \tau}|\phi_i\rangle
\propto \exp{(-S^{(\Delta y)}_{\rm{cl}}[\phi_f,\phi_i])}.
\label{me-from-Scl}
\ee
Here the superscript on $S_{\rm{cl}}$ indicates its dependence on the
width $\Delta y = y_f-y_i$ of the strip.

Now we integrate by parts in the classical action and invoke the boundary
conditions in the $x$ and $y$ direction (for the $x$ direction we use
periodic boundary conditions $\phi_{\rm{cl}}(-L/2,y)=\phi_{\rm{cl}}(L/2,y)$ and then send
the system length $L$ to infinity) as well as the fact that $\phi_{\rm{cl}}(x,y)$ obeys the
Laplace equation. This gives
\be
S^{(\Delta y)}_{\rm{cl}}[\phi_f,\phi_i] = \frac{1}{2\pi K}\int_{-\infty}^{\infty} dx\,
[\phi_f(x)\partial_y \phi(x,y_f) - \phi_i(x)\partial_y \phi(x,y_i)].
\label{Scl-0}
\ee

Next the boundary-value problem for $\phi_{\rm{cl}}(x,y)$ defined on the strip
is mapped to a different geometry where it is more easily solved. Defining the
complex coordinate $z=x+iy$, the conformal transformation
\be
w = \exp{\left(\frac{\pi(z-iy_i)}{\Delta y}\right)}
\label{mapping}
\ee
maps the horizontal strip of width $\Delta y$ in the complex $z$ plane to the upper half of the complex $w$ plane.
In particular, the lower (upper) edge of the strip in the $z$ plane is mapped to the positive
(negative) real axis in the $w$ plane.\footnote{We thank Andrew Doherty for suggesting this mapping.}
Defining $u \equiv \Re(w)$, $v \equiv \Im(w)$, and $\Phi(u,v) \equiv \phi_{\rm{cl}}(x,y)$, we then have
$\Phi(u<0,0)=\phi_f(x)$ and $\Phi(u>0,0)=\phi_i(x)$. Thus $\Phi(u,v)$
is known on the entire real axis $v=0$. Furthermore, it satisfies
the Laplace equation in the upper half plane $v>0$. These properties imply that
$\Phi(u,v)$ is given by the Poisson integral formula \cite{morse}\footnote{Poisson's integral formula is
easily derived using Cauchy's integral formula. See, e.g.,
the derivation of Equation (4.2.13) in \cite{morse}.},
\be
\Phi(u,v) = \frac{v}{\pi}\int_{-\infty}^{\infty} du' \frac{\Phi(u',0)}{(u-u')^2 + v^2}.
\label{poisson-formula}
\ee
To calculate (\ref{Scl-0}) we need $\partial_y \phi(x,y_a) = (\pi/\Delta y)u(x,y_a)
\partial_v \Phi(u(x,y_a),0)$ where
$a=i,f$. From (\ref{poisson-formula}) we have
$\partial_v \Phi(u,0)= - \int_{-\infty}^{\infty} du'\, g(u-u') \Phi(u',0)$, where
\begin{equation}
g(u) \equiv -\lim_{v\to 0}\partial_v\, \delta_v(u).
\label{gx}
\end{equation}
Here $\delta_v(u)$ is a Lorentzian centered at $u=0$ whose width is
determined by $v$,
\begin{equation}
\delta_v(u)=\frac{1}{\pi}\frac{v}{u^2+v^2} =
\frac{1}{2\pi}\int_{-\infty}^{\infty}dq\; e^{-|q|v}e^{iqu}.
\label{lorentzian}
\end{equation}
Note the scaling relation $g(bx)=b^{-2}g(x)$. Using these results we find
\begin{eqnarray}
\fl\lefteqn{S^{(\Delta y)}_{\rm{cl}}[\phi_f,\phi_i] = \frac{1}{2\pi K} \int_{-\infty}^{\infty}dx
\int_{-\infty}^{\infty}dx'}
\nonumber \\ & & \hspace{-1.5cm}\times
\left[\,\sum_{a=i,f}\phi_a(x)\,g(f_{-}(x-x',\Delta y))\,\phi_a(x')
+ 2\, \phi_f(x)\, g(f_{+}(x-x',\Delta y))\,\phi_i(x')\right]
\label{Scl-gen-2}
\end{eqnarray}
where
\be
\fl
f_-(x,\Delta y) = \left(\frac{\pi}{2\Delta y}\right)^{-1}\sinh\left(\frac{\pi x}{2\Delta y}\right),
\qquad
f_+(x,\Delta y) = \left(\frac{\pi}{2\Delta y}\right)^{-1}\cosh\left(\frac{\pi x}{2\Delta y}\right).
\ee
For a generic value of $\Delta y=u\Delta \tau$ the second term in (\ref{Scl-gen-2})
couples $\phi_f$ and $\phi_i$, so that, in agreement with (\ref{me-series}),
$\langle \phi_f|e^{-\hat{H}\Delta \tau}|\phi_i\rangle$ cannot then be written as a product of
two factors with one depending only on $\phi_f$ and the other only on $\phi_i$.
According to (\ref{me-series})-(\ref{gswf-from-me}) such a factorization should
however occur in the limit $\Delta y\to \infty$.
This indeed follows from (\ref{Scl-gen-2}); using that $f_+(x,\Delta y)\to\infty$ and
$f_-(x,\Delta y)\to x$ in this limit, we see that $g(f_+(x-x',\Delta y))\to g(\infty)=0$ [from
$g(x\neq 0)=-1/\pi x^2$] and $g(f_-(x-x',\Delta y))\to g(x-x')$. Thus for $\Delta y\to \infty$ the
second term in (\ref{Scl-gen-2}) vanishes and the classical action reduces to
\be
S^{(\infty)}_{\rm{cl}}[\phi_f,\phi_i] = \frac{1}{2\pi K} \int_{-\infty}^{\infty}dx
\int_{-\infty}^{\infty}dx'\,
\sum_{a=i,f}\phi_a(x)g(x-x')\phi_a(x').
\label{Scl-limit}
\ee
It now follows from (\ref{gswf-from-me}), (\ref{me-from-Scl}) and (\ref{Scl-limit}) that
$\Psi_0[\phi]$ is given by (\ref{gswf-phi}), provided that the function $g(x)$ defined in
(\ref{gx}) is identical to $g(x)$ in (\ref{gswf-phi}). As our notation suggests, this is indeed the
case, as is easily seen from the Fourier representation of the Lorentzian given in
(\ref{lorentzian}), which implies $\tilde{g}(q)=-\lim_{v\to 0}\partial_v\, e^{-|q|v}=|q|$, in agreement
with (\ref{gq}). Thus this path integral derivation gives exactly the same result for $\Psi_0[\phi]$
as the very different derivation using the Schr\"{o}dinger formulation in Sec. \ref{gswf}.

Finally, we note that an analogous derivation of $\bar{\Psi}_0[\theta]=\langle \theta|\Psi_0\rangle$
can be given by considering a path integral representation of $\langle \theta_f|e^{-\hat{H}\Delta \tau}|
\theta_i \rangle$. The only difference from the derivation for $\Psi_0[\phi]$ above is that the Euclidean
action for the $\theta$-field is given by
\begin{equation}
\bar{S}[\theta(x,\tau)] = \frac{K}{2\pi}\int dx \int_{\tau_i}^{\tau_f}
d\tau\; \left(u (\partial_x \theta)^2+\frac{1}{u}(\partial_{\tau}\theta)^2\right).
\label{actiontheta}
\end{equation}
This action displays the same duality as discussed at the end
of Sec. \ref{gswf}, i.e. it can be obtained from the corresponding
action (\ref{actionT=0}) for $\langle \phi_f|e^{-\hat{H}\Delta \tau}|\phi_i\rangle$
by making the replacements $\phi\to \theta$ and $K\to 1/K$. Thus the result
(\ref{gswf-theta}) for $\bar{\Psi}_0[\theta]$ follows.

\end{document}